\documentclass[aps,twocolumn,superscriptaddress]{revtex4-2}
\usepackage[T1]{fontenc}
\usepackage[latin9]{luainputenc}
\usepackage[margin=15mm]{geometry}
\usepackage{float}
\usepackage{amsmath}
\usepackage{amssymb}
\usepackage{graphicx}
\usepackage[colorlinks=true,linkcolor=blue,citecolor=blue,urlcolor=blue]{hyperref}

\usepackage{bbold}
\usepackage{yfonts}
\usepackage{multirow}

\makeatother
\usepackage{babel}

\begin{document}

\title{Inertial spin waves in spin spirals}

\author{Mikhail Cherkasskii}
\email[]{macherkasskii@hotmail.com}
\affiliation{Institute for Theoretical Solid State Physics, RWTH Aachen University, DE-52074 Aachen, Germany}

\author{Ritwik Mondal}
\affiliation{Department of Physics, Indian Institute of Technology (ISM) Dhanbad, IN-826004, Dhanbad, India}

\author{Levente R\'ozsa}
\affiliation{Department of Theoretical Solid State Physics, Institute of Solid State Physics and Optics, HUN-REN Wigner Research Centre for Physics, H-1525 Budapest, Hungary}
\affiliation{Department of Theoretical Physics, Budapest University of Technology and Economics, H-1111 Budapest, Hungary}

\begin{abstract}
Inertial effects in spin dynamics emerge on picosecond time scales, giving rise to nutational excitations at THz frequencies. Here, we describe a general framework for investigating the precessional and nutational excitations in any type of spin structure within linear spin-wave theory. We consider the particular cases of planar and conical spin spirals in detail. We observe a change in the sign of the curvature of the high-frequency nutational spin-wave band as the spiral period is decreased when passing from the ferromagnetic to the antiferromagnetic limit. We identify conditions for the interaction parameters where the curvature changes sign and asymptotical flat bands are formed.
\end{abstract}

\maketitle

\section{Introduction}

Spin waves or magnons are collective excitations in magnetically ordered solids that can propagate through a material without the movement of charge, making them highly promising for information storage and manipulation in low-loss spintronic devices \cite{chumak2015magnon,Hirohata_2014}. The magnonic band structure is possible to detect experimentally using spin-polarized neutron or electron scattering, and Brillouin light scattering~\cite{Vasseur1996,Chisnell2015,Nikotin_1969,Ehrenberg_1999}. The measured band structure can be used to identify the spin structure and to extract microscopic spin-model parameters. 

The competition between different interactions leads to a spatial variation of the spin orientation in so-called non-collinear spin structures. Spin spirals exhibit full rotations of the magnetic moments periodically along one spatial direction, with the spins rotating either in a plane or, often after applying an external field, on the surface of a cone. Such types of magnetic ordering have been observed in a wide array of materials~\cite{Izyumov_1984}, e.g., in rare-earth metals, multiferroics~\cite{Sosnowska_1982}, bulk chiral magnets~\cite{Uchida2006,Bauer2012} or ultrathin magnetic films on heavy-metal substrates~\cite{Bode2007,Yoshida2012,VonBergmann2014}. The excitations of these complex magnetic configurations often exhibit intriguing phenomena like nonreciprocal spin-wave propagation, which have significant implications for spintronic devices~\cite{Garst2017}.

Recent advances in the sub-picosecond manipulation of spins has drawn attention to spin inertia.  
This concept can be described by including an inertial term in the Landau-Lifshitz-Gilbert (LLG) equation, leading to a separation between the directions of the magnetic moment and angular momentum and thereby giving rise to spin nutation~\cite{Olive2012,Olive2015,Bhattacharjee2012,Suhl1998,Bottcher2012,MONDAL2023}. Several theoretical proposals have been set forth to account for the microscopic origin of spin inertia within the LLG equation   \cite{Olive2015,Wegrowe2016JPCM,Mondal2017Nutation,Mondal2018JPCM,Fahnle2011,fahnle2013erratum,Kikuchi}. Theoretical and experimental studies have suggested that the inertial relaxation time 
ranges from a few femtoseconds to several hundred femtoseconds, establishing the characteristic time scale on which nutation can be observed~\cite{Bhattacharjee2012,Bottcher2012,neeraj2019experimental,unikandanunni2021inertial}. The spin inertia results in a nutational resonance in linear response  
that typically falls in the THz regime~\cite{Ciornei2011,Ciornei2010thesis,TitovPRB2021,Cherkasskii2021,cherkasskii2020nutation,Mondal2020nutation}, meaning that it is orders of magnitudes higher than the precessional resonance frequency in ferromagnets. The nutational resonance has been experimenatlly verified in CoFeB and NiFe thin films~\cite{neeraj2019experimental} and in epitaxial Co~\cite{unikandanunni2021inertial}, in which materials an inertial relaxation time of about 300 fs was measured. 

The inertial dynamics gives rise to nutational spin waves propagating in the materials.  
Previous theoretical works on nutational spin waves focused on collinear magnetic structures, i.e., ferromagnets and antiferromagnets  \cite{Kikuchi,Cherkasskii2021,Makhfudz2020,Titov2022PRB,Lomonosov2021,Mondal2022_Inertial_wave}. These investigations predicted that the nutational spin waves appear at a higher frequency, while the inertia decreases the precessional spin-wave frequencies. This is due to the hybridization between the nutation resonance and precession resonance. The nutational spin-wave dispersion follows the familiar parabolic wave-vector dependence of precessional spin waves in ferromagnets, although with a constant frequency shift. In contrast, in two-sublattice collinear antiferromagnets the nutational spin-wave dispersion was found to exhibit a small negative curvature at low wave vectors, i.e., the group velocity points oppositely to the wave vector in this case~\cite{Mondal2022_Inertial_wave}. 

Antiferromagnets can be considered as a periodic spin structure with modulation wave vectors at the edge of the Brillouin zone, while ferromagnets in this description are found at the $\Gamma$ point. Spin spirals with wave vectors along the line connecting these points can be used to interpolate between these two limits, and to shed light on the differences between their nutational spin-wave dispersions. Here, we delve into the dynamics of inertial spin waves  
in spin spirals. First, we derive the equation for linear spin-wave theory in the inertial regime in general non-collinear spin structures. We apply this method to spin spirals to find out at which point the curvature of the nutational spin-wave band is inverted and the band becomes asymptotically flat. Flat bands of single-particle excitations are known to give rise to correlated ordered phases when interactions are taken into account, including flat-band ferromagnetism in fermionic systems~\cite{Tasaki1998}, magnon crystals in quantum spin models~\cite{Zhitomirsky2004}, Wigner crystals in optical lattices of bosons and fermions~\cite{Wu2007}, and topological order in the form of the fractional quantum Hall effect~\cite{sun2011nearly,tang2011high,neupert2011fractional}. It is expected that corrections beyond linear spin-wave theory will significantly influence the inertial spin excitations in such flat bands. 

This paper is organized as follows. In Sec.~\ref{sec:Gen-non-collinear}, we discuss the formalism for inertial spin waves in non-collinear spin structures. In Sec.~\ref{sec:Conical-spiral}, we specify the calculations for a one-dimensional conical spin spiral with nearest-neighbour and next-nearest-neighbour interactions. In Sec.~\ref{sec:Spectra}, we present the calculated spin-wave dispersion relations. In Sec.~\ref{sec:Flat-bands_main}, we derive the conditions for obtaining the flat band. We summarize the results in Sec.~\ref{sec:Conclusion}.

\section{Linear spin-wave theory\label{sec:Gen-non-collinear}}

We adopt the classical Heisenberg model, whose Hamiltonian reads
\begin{equation}
\begin{split}
H & =-\frac{1}{2}\sum_{i,j,\alpha,\beta}{S}_{i}^{\alpha}{J}_{i,j}^{\alpha,\beta}{S}_{j}^{\beta}-\sum_{i,\alpha,\beta}{S}_{i}^{\alpha}{K}_{i,i}^{\alpha,\beta}{S}_{i}^{\beta} \\
&-\sum_{i,\alpha}M_{i}{B}_{i}^{\alpha}{S}_{i}^{\alpha},
\label{eq:Ham_JK}
\end{split}
\end{equation}
where $\alpha$ and $\beta$ are the Cartesian coordinates $\{x,y,z\}$, $\boldsymbol{S}_{i}$ is the spin unit vector at site $i$, ${J}_{i,j}^{\alpha,\beta}$ stands for the elements of the tensorial exchange interaction between sites $i$ and $j$, ${K}_{i,i}^{\alpha,\beta}$ is the single-site anisotropy tensor, $M_{i}$ is the magnitude of the magnetic moment, and $\boldsymbol{ {B}}$ is the external magnetic field. 

The exchange tensor $\boldsymbol{J}_{i,j}$ can be further decomposed as~\cite{Laszloffy2019}
\begin{equation}
\begin{split}
\boldsymbol{J}_{i,j}&=\frac{1}{3}{\rm Tr}\left(\boldsymbol{J}_{i,j}\right)\mathbb{{1}}+\left(\frac{\boldsymbol{J}_{i,j}+\boldsymbol{J}_{i,j}^{T}}{2}-\frac{1}{3}\text{{\rm Tr}}\left(\boldsymbol{J}_{i,j}\right)\mathbb{{1}}\right) \\
&+\frac{\boldsymbol{J}_{i,j}-\boldsymbol{J}_{i,j}^{T}}{2}\nonumber\\
&={\textfrak{J}}_{i,j}+{\textfrak{K}}_{i,j}+{\textfrak{D}}_{i,j}\,.\label{eq:Jcomponents}
\end{split}
\end{equation}
Here, the first term ${\textfrak{J}}_{i,j}$ describes the isotropic exchange, the coefficient of the scalar product $\boldsymbol{S}_{i}\cdot\boldsymbol{S}_{j}$ of the spins in the Hamiltonian. The second term ${\textfrak{K}_{i,j}}$ is the symmetric, traceless two-site anisotropy tensor, which has five independent elements similarly to the single-site anisotropy tensor $\boldsymbol{K}_{i,i}$. The last term  ${\textfrak{D}}_{i,j}$ stands for Dzyaloshinskii--Moriya interaction (DMI)~\cite{Dzyaloshinsky,Moriya}, which may be rewritten as $\boldsymbol{S}_{i}^{T}{\textfrak{D}}_{i,j}\boldsymbol{S}_{j}=\boldsymbol{\mathcal{D}}_{i,j}\left(\boldsymbol{S}_{i}\times\boldsymbol{S}_{j}\right)$ using the Dzyaloshinsky--Moriya vector $\boldsymbol{\mathcal{D}}_{i,j}$.

The dynamics of the spins is described by the inertial Landau-Lifshitz-Gilbert (ILLG) equation, which reads
\begin{equation}
\partial_{t}\boldsymbol{ {S}}_{i}=\boldsymbol{ {{S}}}_{i}\times\left(-\gamma\boldsymbol{ {B}}_{{\rm eff},i}+\alpha{\partial_{t}}\boldsymbol{ \boldsymbol{ {{S}}}}_{i}+\eta{\partial_{tt}}\boldsymbol{ \boldsymbol{ {{S}}}}_{i}\right),\label{eq:ILLG}
\end{equation}
where $\gamma$ is the absolute value of the gyromagnetic ratio, $\alpha$ is the Gilbert damping, $\eta$ is the inertial relaxation time, and the effective magnetic field acting on each spin is defined as
\begin{align}
\boldsymbol{ {B}}_{{\rm eff},i}=-\dfrac{1}{M_{i}}\dfrac{\partial H}{\partial{\boldsymbol{ S}}_{i}}.
\end{align}
If all spins are parallel to the local effective magnetic field,
\begin{equation}
\boldsymbol{ {{S}}}^{(0)}_{i}\times\boldsymbol{ {B}}_{{\rm eff},i}=\boldsymbol{0},\label{eq:ScrossB}
\end{equation}
then the system is in equilibrium, $\partial_{t}\boldsymbol{ {S}}_{i}=\boldsymbol{0}$. Moreover, we only consider equilibrium states which are local minima of the Hamiltonian in Eq.~\eqref{eq:Ham_JK}. To describe non-collinear spin structures, we introduce local coordinate systems, where the quantities will be denoted with a tilde. The local systems at each site are chosen such a way that the equilibrium orientations of the spins are along the \emph{z}-axes, $\tilde{\boldsymbol{S}}_{i}^{(0)}=\boldsymbol{e}_{z}$. Therefore, we define the rotation matrix $\boldsymbol{R}_{i}$ at site \emph{i} as
\begin{align}
\boldsymbol{R}_{i}\tilde{\boldsymbol{S}}_{i}^{(0)}=\boldsymbol{S}_{i}^{(0)}.\label{eq:rotmat}
\end{align}
Introducing the exchange tensor, single-site anisotropy tensor and magnetic field in the local coordinate system as
\begin{align}
\tilde{ \boldsymbol{ B}}_{i} & ={\boldsymbol{ B}}_{i}\boldsymbol{R}_{i},\\
\tilde{\boldsymbol{J}}_{i,j} & =\boldsymbol{R}_{i}^{T}\boldsymbol{J}_{i,j}\boldsymbol{R}_{j},\\
\tilde{\boldsymbol{K}}_{i,i} & =\boldsymbol{R}_{i}^{T}{\boldsymbol{K}}_{i,i}\boldsymbol{R}_{i},
\label{eq:FromLocToGlob}
\end{align}
the Hamiltonian in Eq.~\eqref{eq:Ham_JK} may be rewritten as
\begin{align}
H&=-\frac{1}{2}\sum_{i,j,\alpha,\beta}\tilde{S}_{i}^{\alpha}\tilde{J}_{i,j}^{\alpha,\beta}\tilde{S}_{j}^{\beta}-\sum_{i,\alpha,\beta}\tilde{S}_{i}^{\alpha}\tilde{K}_{i,i}^{\alpha,\beta}\tilde{S}_{i}^{\beta}\\
&-\sum_{i,\alpha}M_{i}\tilde{B}_{i}^{\alpha}\tilde{S}_{i}^{\alpha}.\label{eq:Ham_local}
\end{align}

In linear spin-wave theory, we assume that the spins demonstrate small deviation from the equilibrium direction in the local coordinates, justifying the approximation
\begin{equation}
\tilde{S}_{i}^{z}\approx 1-\frac{1}{2}\left[\left(\tilde{S}_{i}^{x}\right)^{2}+\left(\tilde{S}_{i}^{y}\right)^{2}\right].\label{eq:Sz}
\end{equation}
Substituting Eq.~\eqref{eq:Sz} into Eq.~\eqref{eq:Ham_local} and keeping terms up to second order in the small quantities $\tilde{S}_{i}^{x},\tilde{S}_{i}^{y}$ yields
\begin{align}
H=E_{0}+H_{\textrm{SW}},
\end{align}
where
\begin{align}
E_{0}=-\frac{1}{2}\sum_{i,j}\tilde{J}_{i,j}^{z,z}-\sum_{i}\tilde{K}_{i,i}^{z,z}-\sum_{i}M_{i}\tilde{B}_{i}^{z}
\end{align}
is the energy of the equilibrium state, and
 \begin{align}
 H_{\textrm{SW}}=&  \frac{1}{2}\sum_{i,j}\left[\begin{array}{cc} \tilde{S}_{i}^{x} & \tilde{S}_{i}^{y}\end{array}\right]\left[\begin{array}{cc}A_{1,i,j} & A_{2,i,j} \\ A_{2,j,i} & A_{3,i,j}\end{array}\right]\left[\begin{array}{c} \tilde{S}_{j}^{x} \\ \tilde{S}_{j}^{y}\end{array}\right]
 \end{align}
is the harmonic spin-wave Hamiltonian, with
\begin{align}
A_{1,i,j}=&-\tilde{J}_{i,j}^{x,x}+\delta_{ij}\Bigg(\sum_{k}\tilde{J}_{i,k}^{z,z}-2\tilde{K}_{i,i}^{x,x}\Bigg.\nonumber\\
& + \Bigg. 2\tilde{K}_{i,i}^{z,z}+M_{i}\tilde{B}_{i}^{z}\Bigg),\\
A_{2,i,j}=&-\tilde{J}_{i,j}^{x,y}-\delta_{ij}2\tilde{K}_{i,i}^{x,y},\\
A_{3,i,j}=&-\tilde{J}_{i,j}^{y,y}+\delta_{ij}\Bigg(\sum_{k}\tilde{J}_{i,k}^{z,z}-2\tilde{K}_{i,i}^{y,y}\Bigg.\nonumber\\
& + \Bigg.2\tilde{K}_{i,i}^{z,z}+M_{i}\tilde{B}_{i}^{z}\Bigg).
\end{align}
Note that the linear terms in $\tilde{S}_{i}^{x},\tilde{S}_{i}^{y}$ will vanish due to the equilibrium condition Eq.~\eqref{eq:ScrossB}, and that the spin-wave Hamiltonian is described by a positive-definite matrix around energy minima. Introducing the circularly polarized coordinates $\tilde{S}_{i}^{\pm}=\tilde{S}_{i}^{x}\pm i\tilde{S}_{i}^{y}$, the spin-wave Hamiltonian transforms into
\begin{align}
H_{\textrm{SW}}=&  \frac{1}{4}\sum_{i,j}\left[\begin{array}{cc} \tilde{S}_{i}^{+} & \tilde{S}_{i}^{-}\end{array}\right]\tilde{\mathcal{H}}_{\textrm{SW},i,j}\left[\begin{array}{c} \tilde{S}_{j}^{-} \\ \tilde{S}_{j}^{+}\end{array}\right], \label{eq:H_SW}
\end{align}
where
\begin{align}
\tilde{\mathcal{H}}_{\textrm{SW}}=\frac{1}{2}\left[\begin{array}{cc}\boldsymbol{A}_{1}+\boldsymbol{A}_{3}+i\left(\boldsymbol{A}_{2}-\boldsymbol{A}_{2}^{T}\right) & \boldsymbol{A}_{1}-\boldsymbol{A}_{3}-i\left(\boldsymbol{A}_{2}+\boldsymbol{A}_{2}^{T}\right) \\ \boldsymbol{A}_{1}-\boldsymbol{A}_{3}+i\left(\boldsymbol{A}_{2}+\boldsymbol{A}_{2}^{T}\right) & \boldsymbol{A}_{1}+\boldsymbol{A}_{3}-i\left(\boldsymbol{A}_{2}-\boldsymbol{A}_{2}^{T}\right)\end{array}\right].
\end{align}

Introducing the spin velocities $\tilde{\boldsymbol{V}}_{i}$, the ILLG equation \eqref{eq:ILLG} may be rewritten as a system of first-order differential equations~\cite{Mondal2020nutation},
\begin{align}
\partial_{t}\tilde{\boldsymbol{ {S}}}_{i}= & \tilde{\boldsymbol{ V}}_{i},\\
\partial_{t}\tilde{\boldsymbol{ V}}_{i}= & -\dfrac{1}{\eta}\boldsymbol{ \tilde{S}}_{i}\times\tilde{\boldsymbol{ V}}_{i}-\dfrac{\gamma}{\eta}\tilde{\boldsymbol{ {S}}}_{i}\left(\tilde{\boldsymbol{ {S}}}_{i}\cdot\tilde{\boldsymbol{ B}}_{{\rm eff},i}\right)+\dfrac{\gamma}{\eta} \tilde{\boldsymbol{B}}_{{\rm eff},i}\nonumber\\
 & -\dfrac{\alpha}{\eta}\tilde{\boldsymbol{ V}}_{i}- \tilde{\boldsymbol{ {S}}}_{i}\tilde{\boldsymbol{ V}}_{i}^{2}.
\label{eq:ILLGsys1}
\end{align}
Calculating the effective field $\tilde{\boldsymbol{ B}}_{{\rm eff},i}$ from the spin-wave Hamiltonian Eq.~\eqref{eq:H_SW} and linearizing Eq.~\eqref{eq:ILLGsys1} in the small variables $\tilde{S}^{\pm}_{i}$ and $\tilde{\boldsymbol{V}}_{i}$ results in
\begin{align}
\partial_{t}\left[\begin{array}{c} \tilde{\boldsymbol{S}}^{\perp} \\ \tilde{\boldsymbol{V}}^{\perp}\end{array}\right]=\left[\begin{array}{cc}
0 & \mathbb{{1}}\\
-\gamma\eta^{-1} \boldsymbol{M}^{-1}\tilde{\mathcal{H}}_{\textrm{SW}} & i\eta^{-1}\boldsymbol{\sigma}^{z}-\alpha\eta^{-1}\mathbb{{1}}
\end{array}\right]\left[\begin{array}{c} \tilde{\boldsymbol{S}}^{\perp} \\ \tilde{\boldsymbol{V}}^{\perp}\end{array}\right]\,,\label{eq:matrix_ILLG}
\end{align}
where $\tilde{\boldsymbol{S}}^{\perp}$ is a vector over the lattice sites of components $\tilde{S}_{i}^{\perp}=\left[\tilde{S}_{i}^{-},\tilde{S}_{i}^{+}\right]$, $\tilde{\boldsymbol{V}}^{\perp}$ is defined analogously, $\boldsymbol{M}$ is a diagonal matrix containing the magnetic moments $M_{i}$, and $\boldsymbol{\sigma}^{z}$ is a Pauli matrix in the subspace of the $\pm$ components. Note that in the linear approximation $\partial_{t}\tilde{S}_{i}^{z}=\tilde{V}_{i}^{z}=0$, which is consistent with $\tilde{V}_{i}^{z}$ vanishing from the equations containing four components per lattice site.

The system of homogeneous linear differential equations in Eq.~\eqref{eq:matrix_ILLG} may be solved by assuming a harmonic time dependence, and replacing $\partial_{t}$ by $i\omega$. This yields an eigenvalue equation for the spin-wave frequencies $\omega$ and the eigenmodes. Using the formula for expressing the determinant of $2\times 2$ block matrices with the determinants of the blocks yields the condition
\begin{align}
\textrm{det}\left[-\eta\omega^{2}\mathbb{{1}}+\omega \left(\boldsymbol{\sigma}^{z}+i\alpha\mathbb{{1}}\right)+\gamma \boldsymbol{M}^{-1}\tilde{\mathcal{H}}_{\textrm{SW}}\right]=0\label{eq:secular}
\end{align}
for the eigenvalues. This agrees with the equations derived for the inertial spin-wave modes in ferromagnets and antiferromagnets in Ref.~\cite{Mondal2022_Inertial_wave}. The structure of $\tilde{\mathcal{H}}_{\textrm{SW}}$ ensures that the particle-hole constraint of linear spin-wave theory~\cite{Flynn2020} still holds in the inertial case, meaning that the eigenvalues are found in pairs of $\omega$ and $-\omega^{*}$. In the classical limit discussed here, this simply ensures that the $S_{i}^{x}$ and $S_{i}^{y}$ spin components can always chosen to be real as a linear combination of complex time-dependent solutions $e^{i\omega t}$ and $e^{-i\omega^{*} t}$. Furthermore, taking the limit $\eta\rightarrow 0$ transforms Eq.~\eqref{eq:secular} to
\begin{align}
\textrm{det}\left[\omega \left(\boldsymbol{\sigma}^{z}+i\alpha\mathbb{{1}}\right)+\gamma \boldsymbol{M}^{-1}\tilde{\mathcal{H}}_{\textrm{SW}}\right]=0,\label{eq:secular2}
\end{align}
the well-known generalized eigenvalue equation of spin waves in the non-inertial case~\cite{Garst2017,Rozsa2018}.

\section{Conical spin spiral\label{sec:Conical-spiral}}

\begin{figure}
\begin{centering}
\includegraphics[width=\columnwidth]{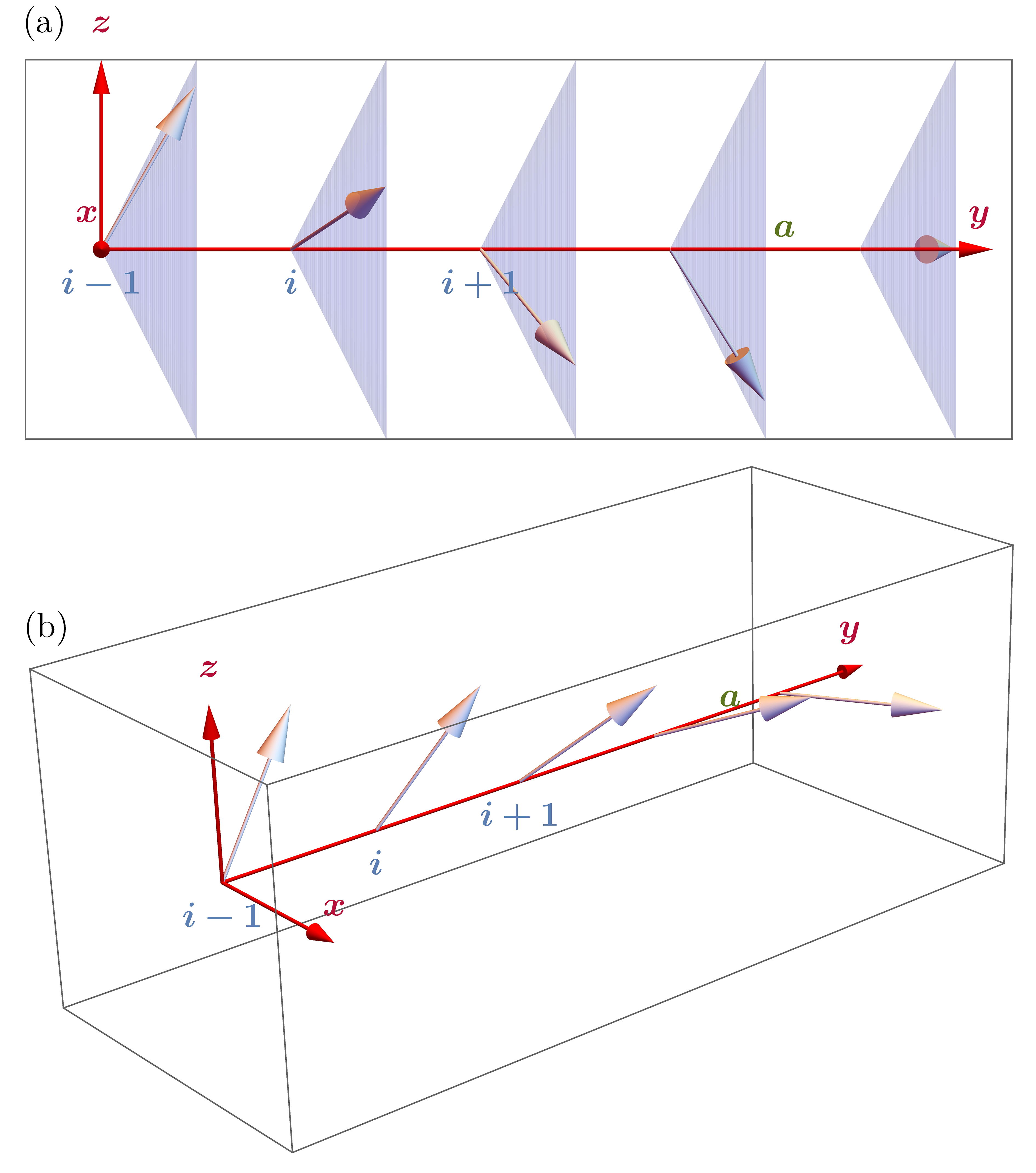}
\par\end{centering}
\caption{\label{fig:conical_spin_spiral}Sketch of a conical spin spiral along the \emph{y}-axis with $S^{y}=0.5$ and the lattice wave vector $\kappa=-\pi/\left(8a\right)$, where $a$ is the lattice constant. Panels (a) and (b) show the same structure, but drawn in different projections. Light blue cones in panel (a) are shown as a guide to the eye.}
\end{figure}

As an application of the general theory, we discuss the spin-wave modes of a conical spin spiral which can be expressed analytically. We consider a one-dimensional chain of atoms along the \emph{y}-axis. In the Hamiltonian Eq.~\eqref{eq:Ham_JK}, we only take into account isotropic exchange interactions $\textfrak{J}_{i,j}=\mathcal{J}_{i,j}\mathbb{{1}}$ as defined in Eq.~\eqref{eq:Jcomponents}, Dzyaloshinsky--Moriya vectors along the \emph{y}-direction $\textfrak{D}_{i,j}^{z,x}=-\textfrak{D}_{i,j}^{x,z}=\mathcal{D}_{i,j}^{y}$, a single-site hard-axis anisotropy along the chain $K_{i,i}^{y,y}=\mathcal{K}>0$, and a magnetic field $B^{y}$ oriented parallel to the chain.

We consider the conical spin spiral
\begin{align}
\boldsymbol{ S}^{(0)}_{i}=
\left[\begin{array}{c}
\sin\left(\kappa r_{i}\right)\sqrt{1-\left(S^{y}\right)^{2}}\\
S^{y}\\
\cos\left(\kappa r_{i}\right)\sqrt{1-\left(S^{y}\right)^{2}}
\end{array}\right],\label{eq:css}
\end{align}
where $r_{i}$ denotes the position along the chain. This structure is illustrated in Fig.~\ref{fig:conical_spin_spiral}. As can be seen in the figure, $\kappa$ determines the period of the harmonic modulation of the spins, and the parameter $S^{y}$ governs the opening angle of the cone. The value $\kappa=0$ describes a ferromagnetic state, while $\kappa=\pi/a,S^{y}=0$ corresponds to the antiferromagnetic state, enabling a continuous transformation between the two limits by changing the wave number $\kappa$. Substituting Eq.~\eqref{eq:css} into the Hamiltonian, the energy of the spiral is found to be
\begin{align}
E=&-\frac{N}{2}\left[\mathcal{J}_{0}\left(S^{y}\right)^{2}+\left(1-\left(S^{y}\right)^{2}\right)\tilde{\mathcal{J}}_{\kappa}\right.\nonumber\\&\left.+2\mathcal{K}\left(S^{y}\right)^{2}+2MB^{y}S^{y}\right],\label{eq:css_energy}
\end{align}
where $N$ is the number of lattice sites, and we introduced the Fourier transforms
\begin{align}
\mathcal{J}_{\kappa}=&\sum_{r_{i}-r_{j}}\mathcal{J}_{i,j}e^{-i\kappa\left(r_{i}-r_{j}\right)}=\sum_{d>0}2\mathcal{J}_{d}\cos\left(\kappa d\right),\\
\mathcal{D}_{\kappa}^{y}=&\sum_{r_{i}-r_{j}}\mathcal{D}_{i,j}^{y}e^{-i\kappa\left(r_{i}-r_{j}\right)}=\sum_{d>0}2i\mathcal{D}_{d}^{y}\sin\left(\kappa d\right),\\
\tilde{\mathcal{J}}_{\kappa}=&\mathcal{J}_{\kappa}+i\mathcal{D}_{\kappa}^{y},
\end{align}
with $d=r_{j}-r_{i}$ the distance between two sites, and we took the symmetry of the isotropic exchange and the antisymmetry of the Dzyaloshinsky--Moriya interaction in site indices into account. Minimizing Eq.~\eqref{eq:css_energy} with respect to the parameters $\kappa$ and $S^{y}$ results in the conditions
\begin{align}
\sum_{d>0}\left(2\mathcal{J}_{d}\sin\left(\kappa d\right)+2\mathcal{D}_{d}^{y}\cos\left(\kappa d\right)\right)=0\label{eq:diffkappa}
\end{align}
and
\begin{align}
S^{y}=\frac{MB^{y}}{\tilde{\mathcal{J}}_{\kappa}-\mathcal{J}_{0}-2\mathcal{K}}.\label{eq:diffSy}
\end{align}
Note that these equations are applicable as long as the right-hand side of Eq.~\eqref{eq:diffSy} is not larger than $1$ in absolute value, otherwise the solution is the collinear state with $S^{y}=1$, and $\kappa$ does not influence the energy. It can be shown that if the parameters minimize the energy, then Eq.~\eqref{eq:css} is also an equilibrium state satisfying Eq.~\eqref{eq:ScrossB}. For further details of the calculation and the generalization to higher-dimensional lattices, see Ref.~\cite{Wuhrer2023}.

The rotation matrices in Eq.~\eqref{eq:rotmat} required for transforming to the local coordinate system read
\begin{equation}
R_{i}=R_{i}^{\left(2\right)}R_{i}^{\text{\ensuremath{\left(1\right)}}},\label{eq:RR}
\end{equation}
with
\begin{equation}
R_{i}^{\text{\ensuremath{\left(1\right)}}}=\left[\begin{array}{ccc}
1 & 0 & 0\\
0 & \sqrt{1-\left(S^{y}\right)^{2}} & S^{y}\\
0 & -S^{y} & \sqrt{1-\left(S^{y}\right)^{2}}
\end{array}\right]\label{eq:R1_Sy}
\end{equation}
a uniform rotation in the \emph{yz}-plane followed by
\begin{equation}
R_{i}^{\left(2\right)}=\left[\begin{array}{ccc}
\cos\left(\kappa r_{i}\right) & 0 & \sin\left(\kappa r_{i}\right)\\
0 & 1 & 0\\
-\sin\left(\kappa r_{i}\right) & 0 & \cos\left(\kappa r_{i}\right)
\end{array}\right],\label{eq:R2}
\end{equation}
a modulated rotation in the $xz$-plane creating the spin spiral. Because the spin spiral is harmonic, the linear spin-wave Hamiltonian in Eq.~\eqref{eq:H_SW} may be brought into a block-diagonal form after Fourier transformation over the atomic sites,
\begin{align}
\tilde{\mathcal{H}}_{\textrm{SW},k_{y}}=\left[\begin{array}{cc}d_{k_{y}} & a^{*}_{-k_{y}} \\ a_{k_{y}} & d^{*}_{-k_{y}}\end{array}\right],
\end{align}
where
\begin{align}
a_{k_{y}}=&-\frac{1}{2}\left[1-\left(S^{y}\right)^{2}\right]\left[\mathcal{J}_{k_{y}}+2\mathcal{K}-\frac{1}{2}\left(\tilde{\mathcal{J}}_{\kappa-k_{y}}+\tilde{\mathcal{J}}_{\kappa+k_{y}}\right)\right],\label{eq:ak}\\
d_{k_{y}}=&a_{k_{y}}+\tilde{\mathcal{J}_{\kappa}}-\frac{1}{2}\left(1+S^{y}\right)\tilde{\mathcal{J}}_{\kappa-k_{y}}-\frac{1}{2}\left(1-S^{y}\right)\tilde{\mathcal{J}}_{\kappa+k_{y}}.\label{eq:dk}
\end{align}
Here we used the energy-minimum condition Eq.~\eqref{eq:diffSy} to eliminate $B^{y}$ from the expressions. Note that $a_{k_{y}}=a^{*}_{-k_{y}}$, but $d_{k_{y}}\neq d^{*}_{-k_{y}}$ if $S^{y}\neq 0$.

Equation~\eqref{eq:secular} determining the magnon modes simplifies to calculating the determinant of $2\times 2$ matrices for each $k$ value. It may be explicitly written as
\begin{align}
&\eta^{2}\omega^{4}-2i\alpha\eta\omega^{3}-\omega^{2}\left(\eta \gamma M^{-1}d_{k_{y}}+\eta \gamma M^{-1}d_{-k_{y}}^{*}+\alpha^{2}+1\right)\nonumber\\&-\omega\gamma M^{-1}\left[\left(1-i\alpha\right)d_{k_{y}}-\left(1+i\alpha\right)d_{-k_{y}}^{*}\right]\nonumber\\&+\gamma^{2} M^{-2}\left(d_{k_{y}}d_{-k_{y}}^{*}-a_{k_{y}}a_{-k_{y}}^{*}\right)=0.\label{eq:disper}
\end{align}
Due to the particle-hole constraint mentioned in Sec.~\ref{sec:Gen-non-collinear}, of the four solutions of the secular equation it is sufficient to treat the two eigenfrequencies with $\textrm{Re}\:\omega>0$. We will denote the solution with the the lower real part of the frequency the precessional spin wave $\omega_{{\rm p}}$, while the higher frequency corresponds to the nutational spin wave $\omega_{{\rm n}}$. In the non-dissipative case $\alpha=0$, these may be expressed as
\begin{equation}
\omega_{{\rm p}}=\dfrac{1}{2}\sqrt{\mathfrak{a}+2W} - \dfrac{1}{2}\sqrt{-3\mathfrak{a}-2W - \frac{2\mathfrak{b}}{\sqrt{\mathfrak{a}+2W}}},\label{eq:wp_4order}
\end{equation}
and
\begin{equation}
\omega_{{\rm n}}=\dfrac{1}{2}\sqrt{\mathfrak{a}+2W} + \dfrac{1}{2}\sqrt{-3\mathfrak{a}-2W - \frac{2\mathfrak{b}}{\sqrt{\mathfrak{a}+2W}}},\label{eq:wn_4order}
\end{equation}
respectively. Here, we used the following notations:
\begin{align}
W= & -\frac{5\mathfrak{a}}{6}+U+V,\\
V= & -\frac{P}{3U},\\
U= & \left[-\frac{Q}{2}-\left(\frac{P^{3}}{27}+\frac{Q^{2}}{4}\right)^{1/2}\right]^{1/3},\\
Q= & \frac{\mathfrak{a}\mathfrak{c}}{3}-\frac{\mathfrak{a}^{3}}{108}-\frac{\mathfrak{b}^{2}}{8},\\
P= & -\frac{\mathfrak{a}}{12}-\mathfrak{c},
\end{align}
\begin{align}
\mathfrak{a}= & -\eta^{-2}\left(\eta \gamma M^{-1}d_{-k_{y}}^{*}+\eta \gamma M^{-1}d_{k_{y}}+1\right),\\
\mathfrak{b}= & \eta^{-2}\gamma M^{-1}\left(d_{-k_{y}}^{*} - d_{k_{y}}\right),\\\mathfrak{c}=& \eta^{-2}\gamma^{2} M^{-2}\left(d_{k_{y}}d_{-k_{y}}^{*}-a_{k_{y}}a_{-k_{y}}^{*}\right).
\end{align}

Since the solutions can be indexed with the wave number $k_{y}$ taking values between $-\pi/a$ and $\pi/a$, for an easier comparison we will show the dispersion relation in this atomic or extended Brillouin zone~\cite{Garst2017} for all values of the spiral wave number $\kappa$. Note that extending the Hamiltonian by additional terms may introduce hybridizations between the modes and open gaps in the spectrum at integer multiples of $\kappa/2$, in which case the band structure would be more conveniently visualized via multiple bands inside the magnetic Brillouin zone between $-\kappa/2$ and $\kappa/2$.

\section{Spin-wave dispersions\label{sec:Spectra}}

\begin{figure*}
\begin{centering}
\includegraphics[width=2\columnwidth]{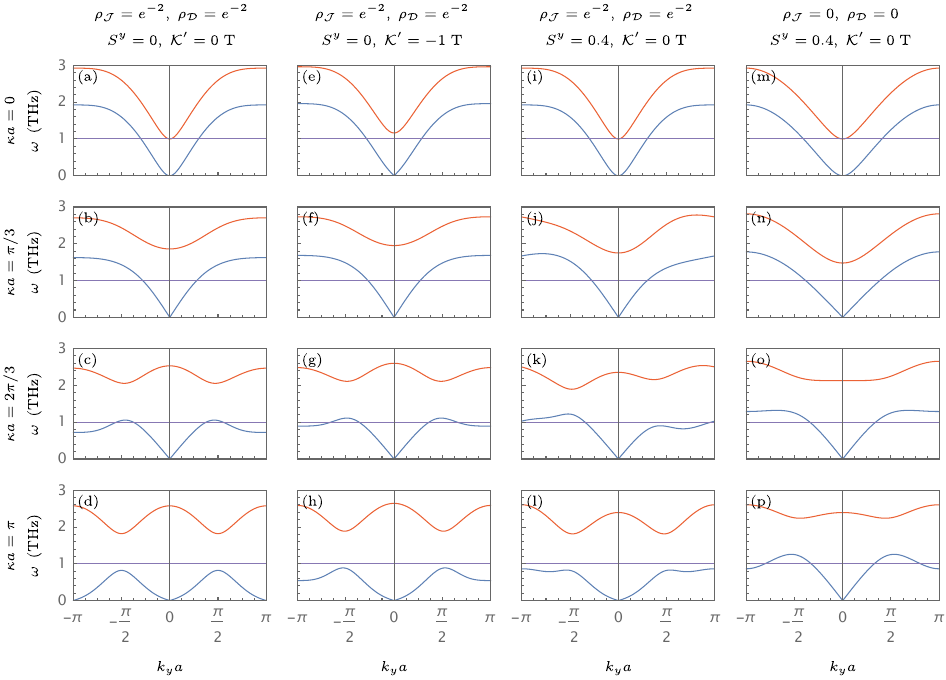}
\par\end{centering}
\caption{\label{fig:disp_num}Dispersion relations of precession (blue) and nutation (red) spin waves in (a)-(d) planar spiral with next-nearest-neighbor coupling for $\mathcal{K}'=0$~T, (e)-(h) planar spiral with next-nearest-neighbor coupling for $\mathcal{K}'=-1$~T, (i)-(l) conical spiral with next-nearest-neighbor coupling for $\mathcal{K}'=0$~T, and (m)-(p) conical spiral with only nearest-neighbor coupling for $\mathcal{K}'=0$~T. The period of the spiral is varied between the rows of panels. The purple line shows the characteristic inertial frequency $\eta^{-1}$. 
We used the parameters: $\gamma=2\pi\times28\;\text{GHz/T}$, $\alpha=0$, $\eta=1\;\text{ps}$, $\mathfrak{\mathcal{E}'=8\;\text{T}}$ and the varied values listed in Table~\ref{tab:param}. }
\end{figure*}

\begin{table*}
\begin{centering}
\begin{tabular}{|l|r|r|l|r|r|l|r|r|l|r|r|}
\hline 
\hline 
\multicolumn{3}{|c|}{$\rho_{\mathcal{J}}=e^{-2},\;\rho_{\mathcal{D}}=e^{-2}$} & \multicolumn{3}{c|}{$\rho_{\mathcal{J}}=e^{-2},\;\rho_{\mathcal{D}}=e^{-2}$} & \multicolumn{3}{c|}{$\rho_{\mathcal{J}}=e^{-2},\;\rho_{\mathcal{D}}=e^{-2}$} & \multicolumn{3}{c|}{$\rho_{\mathcal{J}}=0,\;\rho_{\mathcal{D}}=0$}\tabularnewline
\hline 
\multicolumn{3}{|c|}{$S^{y}=0,\;\mathcal{K}'=0$~T} & \multicolumn{3}{c|}{$S^{y}=0,\;\mathcal{K}'=-1\;\text{T}$} & \multicolumn{3}{c|}{$S^{y}=0.4,\;\mathcal{K}'=0$~T} & \multicolumn{3}{c|}{$S^{y}=0.4,\;\mathcal{K}'=0$~T}\tabularnewline
\hline 
$\kappa a$ & $\mathcal{J}'\;\left(\text{T}\right)$ & $\mathcal{D}'{}^{y}\;\left(\text{T}\right)$ & $\kappa a$ & $\mathcal{J}'\;\left(\text{T}\right)$ & $\mathcal{D}'{}^{y}\;\left(\text{T}\right)$ & $\kappa a$ & $\mathcal{J}'\;\left(\text{T}\right)$ & $\mathcal{D}'{}^{y}\;\left(\text{T}\right)$ & $\kappa a$ & $\mathcal{J}'\;\left(\text{T}\right)$ & $\mathcal{D}'{}^{y}\;\left(\text{T}\right)$\tabularnewline
\hline 
\hline 
(a) $0$ & $8.00$ & $0.00$ & (e) $0$ & $8.00$ & $0.00$ & (i) $0$ & $8.00$ & $0.00$ & (m) $0$ & $8.00$ & $0.00$\tabularnewline
\hline 
(b) $\pi/3$ & $1.63$ & $7.83$ & (f) $\pi/3$ & $1.63$ & $7.83$ & (j) $\pi/3$ & $1.63$ & $7.83$ & (n) $\pi/3$ & $4$ & $6.93$\tabularnewline
\hline 
(c) $2\pi/3$ & $-6.78$ & $4.24$ & (g) $2\pi/3$ & $-6.78$ & $4.24$ & (k) $2\pi/3$ & $-6.78$ & $4.24$ & (o) $2\pi/3$ & $-4$ & $6.93$\tabularnewline
\hline 
(d) $\pi$ & $-8.00$ & $0.00$ & (h) $\pi$ & $-8.00$ & $0.00$ & (l) $\pi$ & $-8.00$ & $0.00$ & (p) $\pi$ & $-8.00$ & $0.00$\tabularnewline
\hline 
\end{tabular}
\par\end{centering}
\caption{\label{tab:param}Parameters used for the numerical calculations shown in Fig.~\ref{fig:disp_num}.}
\end{table*}

We numerically investigated the dependence of the precession and nutation bands on the spin-spiral period $\kappa$ and the opening angle of the cone characterized by $S^{y}$. We restricted the interactions to nearest and next-nearest neighbors as $\mathcal{J}_{1}=M\mathcal{J}'$, $\mathcal{J}_{2}=M\rho_{\mathcal{J}}\mathcal{J}'$, $\mathcal{D}^{y}_{1}=M\mathcal{D}'{}^{y}$, $\mathcal{D}^{y}_{2}=M\rho_{\mathcal{D}}\mathcal{D}'{}^{y}$ and $\mathcal{K}=M\mathcal{K}'$, where the interaction coefficients denoted with a prime are in units of magnetic field and $\rho_{\mathcal{J}},\rho_{\mathcal{D}}$ are dimensionless ratios of the interactions between second and first neighbors. We selected the values $\rho_{\mathcal{J}}$, $\rho_{\mathcal{D}}$, and $\mathcal{K}'$. In order to keep the total frequency range of the dispersion the same as the period is varied, we also fixed the parameter
\begin{equation}
\mathcal{E}'^{2}=\mathcal{J}'^{2}+\left(\mathcal{D}'{}^{y}\right)^{2}.\label{eq:J2plusD2}
\end{equation}
We chose the $\kappa$ and $S^{y}$ values, and determined the spin-model parameters $\mathcal{J}'$, $\mathcal{D}'{}^{y}$, and $B^{y}$ based on the conditions Eq.~\eqref{eq:diffkappa} and \eqref{eq:diffSy}, i.e., we determined the Hamiltonian such a way that it is minimized for the selected period and cone angle. In particular, $\mathcal{J}'$ and $\mathcal{D}'{}^{y}$ were expressed from Eq.~\eqref{eq:J2plusD2} and
\begin{align}
\frac{\mathcal{D}'{}^{y}}{\mathcal{J}'}= -\frac{\sin\left(\kappa a\right)+2\rho_{\mathcal{J}}\sin\left(2\kappa a\right)}{\cos\left(\kappa a\right)+2\rho_{\mathcal{D}}\cos\left(2\kappa a\right)}.
\label{eq:J2andD2}
\end{align}

The chosen parameter values are summarized in Table~\ref{tab:param} and the numerically calculated spectra are shown in Fig.~\ref{fig:disp_num}. Note that this procedure of parameter choice is not meant to describe a specific material, since the magnetic interaction parameters vary even in orders of magnitude between different systems hosting spin spirals, while at the moment there is rather limited information about the values of the inertial relaxation times. We intend to illustrate the possible qualitative effects of inertial dynamics on the magnon band structure as the period and the cone angle of the spiral is varied, while keeping the range of the wave vector and frequencies fixed to facilitate an easier comparison between the different cases. 
Figure~\ref{fig:disp_num}(a) displays the familiar ungapped parabolic precessional dispersion relation of the ferromagnetic state for nearest-neighbor and next-nearest-neighbor couplings. The nutational band is shifted by a constant $\eta^{-1}$ to higher frequencies~\cite{Mondal2022_Inertial_wave}. The spectrum is identical when the ferromagnetic direction is rotated out from the \emph{z}-axis towards the \emph{y}-axis (Fig.~\ref{fig:disp_num}(i)) since the system is invariant under global spin rotations. The dispersion slightly changes quantitatively when the next-nearest-neighbor interactions are turned off (Fig.~\ref{fig:disp_num}(m)). A qualitative difference can only be observed if a hard-axis anisotropy $\mathcal{K}'$ is introduced perpendicular to the plane (Fig.~\ref{fig:disp_num}(e)): this term cannot gap out the Goldstone mode of the precessional band, since the system remains rotationally invariant around the \emph{y}-axis, but the nutational frequency at zero wave vector is slightly increased compared to $\eta^{-1}$, and the two branches are no longer shifted by only a constant value with respect to each other.

The dispersion relations for the $\kappa a=\pi$ two-sublattice structures are displayed in the bottom row of Fig.~\ref{fig:disp_num}. In the isotropic antiferromagnet (Fig.~\ref{fig:disp_num}(d)), the dispersion is periodic with $k_{y} a=\pi$ in the extended Brillouin zone; this would correspond to a double-degenerate band when the dispersion relation is folded back into the magnetic Brillouin zone between $-\pi/(2a)$ and $\pi/(2a)$. Note that while the precession band is linear around $k_{y}a=0$ with a finite group velocity, the nutational band is parabolic with a negative curvature~\cite{Mondal2022_Inertial_wave}. The hard-axis anisotropy perpendicular to the spins (Fig.~\ref{fig:disp_num}(h)) does not lift the Goldstone mode at zero wave vector, but it introduces a gap at $k_{y}a=\pi$, meaning that the spectrum would no longer be degenerate in the magnetic Brillouin zone. The nutational band is less affected by the anisotropy. Tilting the spins towards the \emph{y}-axis by an external magnetic field (Fig.~\ref{fig:disp_num}(l) and (p)) decreases the angle between the two sublattices, leading to a spin-flop state. While the Goldstone mode is preserved, this also increases the frequencies at higher wave vectors, since the ferromagnetic limit has to be reached at $S^{y}=1$. This already leads to a local maximum at $k_{y}a=\pi$ in Fig.~\ref{fig:disp_num}(l). The reduced cone angle considerably flattens the nutational band around $k_{y}a=0$, particularly in Fig.~\ref{fig:disp_num}(p). This can again be understood by approaching the ferromagnetic configuration, since at that point the curvature of the dispersion has to revert back to positive.

The transformation between the ferromagnetic and the antiferromagnetic state in the dispersion can also be observed when changing the spiral wave vector $\kappa$ in Fig.~\ref{fig:disp_num}(b), (c), (f), (g), (j), (k), (n), and (o). The Goldstone mode associated with a global shift of the phase of the spiral is observable in all panels. The largest influence of increasing the wave vector can be seen in the nutational band around $k_{y}a=0$ where the minimum is turned into a local maximum, and in the precessional band around $k_{y}a=\pi$ where the maximum is turned into a minimum. Because the inversion of the curvature happens at different wave vectors in the precessional and nutational bands, observing such an inversion could distinguish between inertial and non-inertial effects in the dynamics. The effect of the hard-axis anisotropy is best visible in increasing the precessional frequencies at $k_{y}a=\pi$ (Fig.~\ref{fig:disp_num}(f) and (g)). Increasing the collinear component $S^{y}$ is qualitatively similar to decreasing the wave vector, since both decrease the angle between the neighboring spins. In Fig.~\ref{fig:disp_num}(j) and (k), a non-reciprocity can be observed in the dispersion relation, i.e., the frequencies between positive and negative $k_{y}$ values differ. Note that the preferred propagation direction for spin waves, i.e., the lower value of the frequency between $\omega(k_{y})$ and $\omega(-k_{y})$, is opposite between precessional and nutational waves, similarly to the case of collinear configurations discussed in Ref.~\cite{Mondal2022_Inertial_wave}. In the considered system, the non-reciprocity has three necessary criteria: (i) a finite spiral wave vector $\kappa a\notin\{0,\pi\}$, (ii) a finite collinear component $S^{y}$, and (iii) a finite next-nearest-neighbour interaction $\rho_{\mathcal{J}}$ (cf. Fig.~\ref{fig:disp_num}(n) and (o) where the non-reciprocity vanishes). Note that (ii) is generally necessary for breaking the effective time-reversal symmetry of the spirals that protects the reciprocity of the dispersion. However, condition (i) may be weakened and the non-reciprocity observed in ferromagnetic states when the Dzyaloshinsky--Moriya interaction is pointing along the ferromagnetic orientation in a different geometry~\cite{Udvardi2009}. Condition (iii) is only required because of the special mathematical structure of the nearest-neighbor model~\cite{Wuhrer2023}.

\begin{figure*}
\begin{centering}
\includegraphics[width=2\columnwidth]{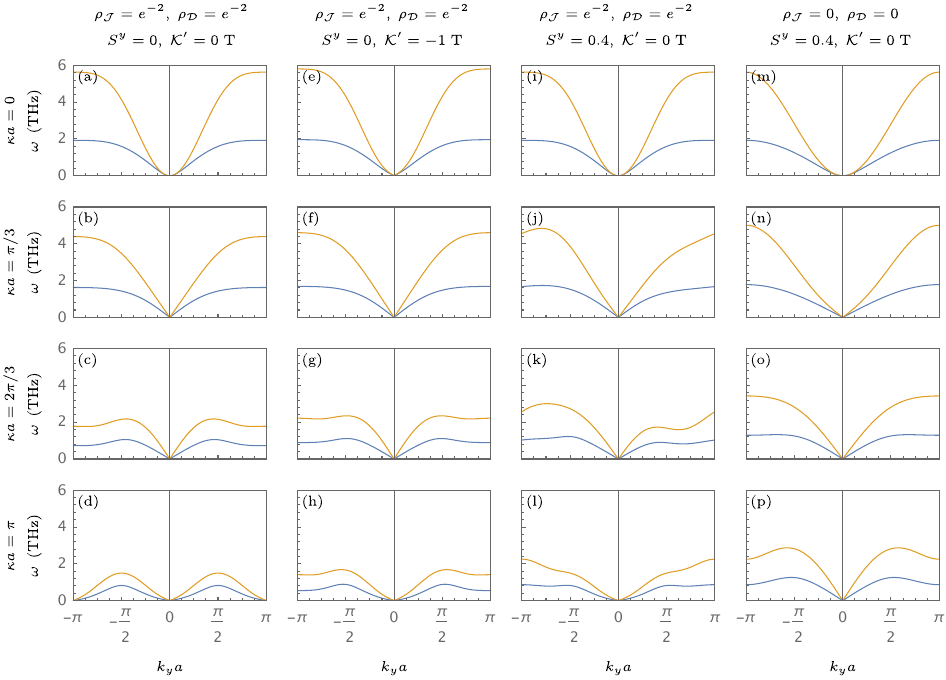}
\par\end{centering}
\caption{\label{fig:NoInert_disp}Dispersion relations of inertial (blue curves at $\eta=1\;\text{ps}$) and non-inertial (orange curves at $\eta=0$~ps) precessional spin waves. The nutational bands are not shown. Panels (a)-(d) show a planar spiral with next-nearest-neighbor coupling for $\mathcal{K}'=0$~T, (e)-(h) a planar spiral with next-nearest-neighbor coupling for $\mathcal{K}'=-1$~T, (i)-(l) a conical spiral with next-nearest-neighbor coupling for $\mathcal{K}'=0$~T, and (m)-(p) a conical spiral with only nearest-neighbor coupling for $\mathcal{K}'=0$~T. The period of the spiral is varied between the rows of panels. The parameters are the same as in Fig.~\ref{fig:disp_num}: $\gamma=2\pi\times28\;\text{GHz/T}$, $\alpha=0,$ $\mathfrak{\mathcal{E}'=8\;\text{T}}$, and the values given in Table~\ref{tab:param}. }
\end{figure*}

We present a comparison between the dispersion characteristics of precessional spin waves both in the presence and the absence of inertia in Fig.~\ref{fig:NoInert_disp}. The frequencies are decreased by the presence of inertia in all cases. The ratio of the frequencies in the presence and in the absence of inertia was found in Ref.~\cite{Mondal2022_Inertial_wave} to show a pronounced dependence on the wave vector in ferromagnets, while in antiferromagnets it is almost independent of the wave vector. In spin spirals, the dependence on the wave vector is most obvious in the case of the non-reciprocal dispersions (Fig.~\ref{fig:NoInert_disp}(j) and (k)), where the positions of the minima and maxima are also shifted. However, it is not possible to directly compare the precessional dispersions with and without the inertial term in experiments, and the ``bare'' values of the interaction parameters determining the shape of the dispersion are not known, either. Therefore, the most convincing experimental evidence of inertial dynamics would have to rely on the detection of the nutational spin waves.

\section{Condition for the flat band\label{sec:Flat-bands_main}}

We observed in Sec.~\ref{sec:Spectra} that the curvature of the nutational band at the center of the Brillouin zone has different signs in the ferromagnetic and the antiferromagnetic limits, and the sign may be inverted by changing the wave vector of the spiral or the cone angle. At the point where the curvature vanishes, the nutational band becomes relatively flat in a wide range of wave vectors. We will derive a condition on the model parameters where this inversion of the curvature occurs. We will consider the case $S^{y}=0$, because then the dispersion is reciprocal, meaning that there is always a maximum or a minimum at $k_{y}a=0$. In this case, the matrix elements of the spin-wave Hamiltonian in Eqs.~\eqref{eq:ak} and \eqref{eq:dk} simplify to
\begin{align}
a_{k_{y}}=&\frac{1}{2}\left[\frac{1}{2}\left(\tilde{\mathcal{J}}_{\kappa-k_{y}}+\tilde{\mathcal{J}}_{\kappa+k_{y}}\right)-\mathcal{J}_{k_{y}}-2\mathcal{K}\right]=a^{*}_{-k_{y}},\label{eq:ak2}\\
d_{k_{y}}=&\frac{1}{2}\left[2\tilde{\mathcal{J}}_{\kappa}-\frac{1}{2}\left(\tilde{\mathcal{J}}_{\kappa-k_{y}}+\tilde{\mathcal{J}}_{\kappa+k_{y}}\right)-\mathcal{J}_{k_{y}}-2\mathcal{K}\right]=d^{*}_{-k_{y}},\label{eq:dk2}
\end{align}
and the spin-wave frequencies may be written in a closed form as
\begin{align}
\begin{split}
& \omega= \left\{ \dfrac{1}{2\eta^{2}}\left[1+2\eta\gamma M^{-1}d_{k_{y}}\right]\right.\\
 & \left.\pm\dfrac{1}{2\eta^{2}}\left[1+4\eta\gamma M^{-1}d_{k_{y}}+4\left(\eta\gamma M^{-1}a_{k_{y}}\right)^{2}\right]^{\frac{1}{2}}\right\} ^{\frac{1}{2}},
\end{split}
\end{align}
where the negative and positive signs pertain to precessional and nutational bands, respectively. The point where the curvature inverts can be obtained by expanding the nutational frequency in $k_{y}$ around $k_{y}a=0$, and determining where the quadratic term vanishes. Note that the linear and cubic terms in $k_{y}$ vanish due to the reciprocity of the dispersion, leaving the quartic terms as the leading correction at this point. This yields the condition
\begin{align}
\begin{split}
&\eta\gamma M^{-1}\left(\partial_{k_{y}}^{2}\tilde{\mathcal{J}}_{\kappa}+\partial_{k_{y}}^{2}\mathcal{J}_{0}\right)\\
&+\left(\eta\gamma M^{-1}\right)^{2}\left(\tilde{\mathcal{J}}_{\kappa}-\mathcal{J}_{0}-2\mathcal{K}\right)\partial_{k_{y}}^{2}\mathcal{J}_{0}=0.
\end{split}
\end{align}
Considering only nearest-neighbor interactions and using the parameter $\mathcal{E}'$ from Eq.~\eqref{eq:J2plusD2}, the wave vector $\kappa$ for which this condition is satisfied may be given in a closed form:
\begin{align}
\begin{split}
\cos\left(\kappa a\right)&=
\dfrac{1}{4\eta\gamma\mathcal{E}'}\left[1+2\eta\gamma\left(\mathcal{E}'-\mathcal{K}'\right)\right]\\
&-\dfrac{1}{4\eta\gamma\mathcal{E}'}\sqrt{\left[1+2\eta\gamma\left(\mathcal{E}'-\mathcal{K}'\right)\right]^{2}+8\eta\gamma\mathcal{E}'}.\label{eq:flatbandcond}
\end{split}
\end{align}
Note that $\mathcal{E}'>0$ and $\mathcal{K}'<0$, meaning that $\cos\left(\kappa a\right)<0$, i.e., the inversion of the curvature happens between $\kappa a=\pi/2$ and $\kappa a=\pi$.

\begin{figure}
\begin{centering}
\includegraphics[width=\columnwidth]{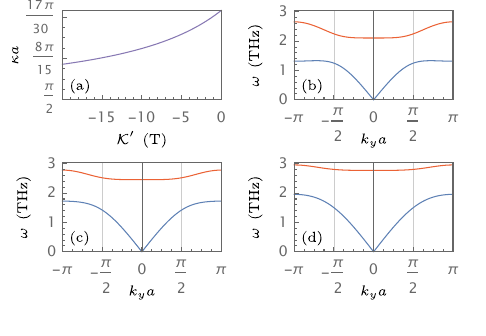}
\par\end{centering}
\caption{\label{fig:Kyy_vs_kappa_a}(a) The relation between single-site anisotropy $\mathcal{K}'$ and wave vector $\kappa$ given by Eq.~\eqref{eq:flatbandcond}, at which the curvature of the nutational band vanishes. (b-d) display the precessional (blue) and nutational (red) bands for the parameters (b) $\mathcal{K}'=0$~T, $\mathcal{J}'=-1.79\;\text{T}$, $\mathcal{D}'{}^{y}=7.79\;\text{T}$, $\kappa a=\pi/2+0.07\pi$; (c) $\mathcal{K}'=-5\;\text{T}$, $\mathcal{J}'=-1.32\;\text{T}$, $\mathcal{D}'{}^{y}=7.88\;\text{T}$, $\kappa a=\pi/2+0.052\pi$; and (d) $\mathcal{K}'=-10\;\text{T}$, $\mathcal{J}'=-1.04\;\text{T}$, $\mathcal{D}'{}^{y}=7.93\;\text{T}$, $\kappa a=\pi/2+0.041\pi$. The other parameters are $\mathcal{E}'=8\;\text{T}$, $\gamma=2\pi\times28\;\text{GHz/T}$, $\alpha=0$, $\eta=1\;\text{ps}$, $S^{y}=0$, $\rho_{\mathcal{J}}=0$, $\rho_{\mathcal{D}}=0$ in all panels. }
\end{figure}

Based on Eq.~\eqref{eq:flatbandcond}, we illustrate the relation between $\mathcal{K}'$ and $\kappa a$ in Fig.~\ref{fig:Kyy_vs_kappa_a}(a). Increasing the magnitude of the anisotropy $\mathcal{K}'$ shifts $\kappa a$ closer to $\pi/2$. It can be seen from Eq.~\eqref{eq:flatbandcond} that $\kappa a$ approaches $\pi$, i.e., the antiferromagnetic configuration, as the inertial parameter $\eta$ is decreased. As can be seen in Fig.~\ref{fig:Kyy_vs_kappa_a}(b)-(d), at the point where the condition is satisfied the nutational band appears flat in a rather large range approximately between $-\pi/2<k_{y}a<\pi/2$. The total width of the nutational band decreases as the magnitude of $\mathcal{K}'$ is increased. Note that the precessional band appears flat in the vicinity of the point $k_{y}a=\pi$, where the curvature also inverts as mentioned in Sec.~\ref{sec:Spectra}. However, the inversion of the curvature of the precessional band happens at a numerically different value than that of the nutational band, and the precessional band becomes less flat at the point where Eq.~\eqref{eq:flatbandcond} is satisfied as the magnitude of $\mathcal{K}'$ is increased.

\section{Conclusion\label{sec:Conclusion}}

In this study, we discussed precessional and nutational spin waves in non-collinear spin configurations based on the linearization of the inertial Landau--Lifshitz--Gilbert equation. We derived the general formula for determining the spin-wave frequencies and eigenmodes, and discussed how it transforms in the inertial-free limit. We applied the method to calculate the dispersion relations for conical spin spirals in one dimension. We found the requirements for observing a non-reciprocal dispersion in the considered model, and observed that the curvature of the nutational band is inverted around $k_{y}a=0$ when passing from the ferromagnetic to the antiferromagnetic limit, while in the precessional band the inversion occurs around $k_{y}a=\pi$ in the extended Brillouin zone. We derived the condition for the inversion of the curvature of the nutational band, and demonstrated that the band becomes relatively flat in a wide range of wave vectors.

Although the observed nutational bands are not flat in the whole extended Brillouin zone, the nutational band will split into multiple bands at the boundary of the magnetic Brillouin zone at $\pm\kappa/2$ when the spin spiral becomes anharmonic and the spin waves hybridize with each other. Such flat bands in reduced magnetic Brillouin zones, also known as spin-wave Landau levels, have been investigated in detail in non-collinear spin structures in the non-inertial limit in Ref.~\cite{Weber2022}. Based on the analogy with electronic systems, it is also expected that due to the high density of states in the flat bands, linear spin-wave theory may no longer accurately describe the nature of the excitations, and the interaction between the spin waves plays a more pronounced role. The formalism presented here may stimulate further theoretical and experimental investigations of inertial spin waves in various kinds of non-collinear spin structures.

\textit{Note:} The general formulae for the spin-wave eigenmodes of the inertial Landau--Lifshitz--Gilbert equation has been recently derived in Ref.~\cite{Dannegger} independently of our work.

\section*{Acknowledgments}

We are grateful to Ulrich Nowak, Tobias Dannegger and David Angster for fruitful discussions. Financial support by the faculty research scheme at IIT (ISM) Dhanbad, India under Project No. FRS(196)/2023-2024/PHYSICS, by the Science and Engineering Research Board (SERB), India under Project No. SRG/2023/000612, by the National Research, Development, and Innovation Office (NRDI) of Hungary under Project Nos. K131938 and FK142601, by the Ministry of Culture and Innovation and the National Research, Development and Innovation Office within the Quantum Information National Laboratory of Hungary (Grant No. 2022-2.1.1-NL-2022-00004), by the Hungarian Academy of Sciences via a J\'{a}nos Bolyai Research Grant (Grant No. BO/00178/23/11) are gratefully acknowledged.

\end{document}